
\magnification=\magstep1
\hsize=13.5cm
\vsize=19cm
\overfullrule 0pt
\baselineskip=10pt plus1pt minus1pt
\lineskip=3.5pt plus1pt minus1pt
\lineskiplimit=3.5pt
\parskip=4pt plus1pt minus4pt
\leftskip=0.4cm

\def\negenspace{\kern-1.1em}


\newcount\secno
\secno=0
\newcount\susecno
\newcount\fmno\def\z{\global\advance\fmno by 1 \the\secno.
            \the\susecno.\the\fmno}
\def\section#1{\global\advance\secno by 1
    \susecno=0 \fmno=0
    \centerline{\bf \the\secno. #1}\par}
\def\subsection#1{\medbreak\global\advance\susecno by 1
      \fmno=0
  \noindent{\the\secno.\the\susecno. {\it #1}}\noindent}


\newcount\refno
\refno=1
\def\y{\the\refno}
\def\myfoot#1{\footnote{$^{(\y)}$}{#1}
    \advance\refno by 1}


\def\neq{\hbox{$\,$=\kern-6.5pt /$\,$}}







\newcount\secno
\secno=0
\newcount\fmno\def\z{\global\advance\fmno by 1 \the\secno.
    \the\fmno}
\def\sectio#1{\medbreak\global\advance\secno by 1
      \fmno=0
  \noindent{\the\secno. {\it #1}}\noindent}

\def\Reel{R}

\def\eqa{\eqno(\z )}\def\vta{\vartheta}

\def\a{\alpha}
\def\b{\beta}
\def\c{\gamma}
\def\d{\delta}
\def\vta{\vartheta}
\def\eps{\varepsilon}
\def\t{\theta}
\def\p{\varphi}
\def\o{\Omega}

\rightline{CPP-94-38}
\bigskip
\centerline{\bf ON NON-RIEMANNIAN PARALLEL TRANSPORT}
 \centerline{\bf IN REGGE CALCULUS}
\bigskip
\centerline{by}
\bigskip
\centerline{Frank Gronwald$^{*}$\myfoot{permanent address:
Institute for Theoretical Physics,
University of Cologne, D-50923 K\"oln, Germany.  e-mail: fg@thp.uni-koeln.de}}

\bigskip
\noindent $^{*})$ Center for Particle Physics, Department of Physics,
University of Texas, Austin, Texas 78712
\bigskip
\centerline{\bf Abstract}
\bigskip
We discuss the possibility of incorporating non-Riemannian parallel
transport into Regge calculus. It is shown that every Regge lattice is
locally equivalent to a space of constant curvature. Therefore well
known-concepts of differential geometry imply the definition of an arbitrary
linear affine connection on a Regge lattice.
\bigskip
\noindent
PACS-numbers: 02.40, 04.60

\vfill
\eject

\sectio{\bf Introduction}

The interest in Regge calculus [1] and its use as an
approximation scheme for Riemannian manifolds has increased
during the last fifteen years. Regge calculus was applied to a variety of
problems in both classical
and quantum gravity, see [2] for a recent review.
Regge calculus requires the approximation of smooth manifolds
by piecewise flat spaces which are built from flat simplexes. The
curvature of such an $n-$dimensional piecewise flat space resides in its
$(n-2)$--dimensional subsimplexes, usually referred to as the ``hinges''
of the ``Regge lattice''.

Regge calculus is a useful tool for numerical calculations in Riemannian
spacetimes. It is unsatisfactory that it cannot be applied
to non-Riemannian manifolds so far. Possible applications
of a non-Riemannian Regge calculus would include gauge theories of
gravity [3] and supergravity in ordinary spacetime [4]. These Yang-Mills type
extensions of general relativity are especially important when it comes
to the inclusion of fermionic matter fields. Fermionic
matter fields are known to be possible sources of torsion.
Their introduction requires a linear connection that a priori
is independent of the metric, i.e., one has to use non-Riemannian geometries
as an appropriate framework. But an enlargement of Regge calculus to
non-Riemannian manifolds is also of general interest. It should lead to a
better understanding of the Regge lattice, showing more clearly its relation
to ordinary differential geometry.

In order to extend Regge calculus to the
non-Riemannian case, it is inevitable to
investigate the notion of a linear connection in a Regge lattice.
The metric of the Regge lattice is given
by its link lengths. In order to implement a non-Riemannian connection,
it is usually proposed to let a parallelly transported frame undergo a
non--trivial rotation [5--8], a linear transformation [9], or a
Poincar{\'e} transformation
[10] while it is passing from one $n-$simplex to another.
It was argued in [6]  that the non--trivial rotations do not
contribute to the Einstein--Cartan action. This is not quite correct
because the contributions from the non-trivial rotations to the curvature
associated with the hinges were not properly calculated. In fact, the
nontrivial rotations can be represented by a discrete version of the
contortion, i.e. the non-Riemannian part of the linear connection.
A non--trivial contortion will generally produce torsion and a non-Riemannian
piece to the curvature.

It is the purpose of this paper to clarify the geometric content of an
arbitrary Regge lattice of dimension two, three, and four.
We show in Sec.2 that the geometry of such a Regge lattice is locally
equivalent to the geometry of a specific space of constant curvature.
This is done by an isometric embedding of the basic building block of
Regge calculus, the $(\eps -n)-$cone, into $\Reel^{n+1}$. Parallel transport
on this embedded cone is, in turn, seen to be equivalent to parallel transport
in a space of constant curvature. This, together with a short analysis of
Cartan's structure equations, reveals in Sec.3 how a general parallel
transport should be defined in a Regge lattice. It is shown how non-Riemannian
quantities appear by parallel transport around hinges. We summarize the
results in Sec.4.

\bigskip
\goodbreak
\sectio{\bf Isometric embedding of Regge's $(\eps -n)-$cone for $n=2,3,4$}

\bigskip
\noindent
{\bf 2.1. Preliminaries}

The basic building block of Regge calculus is Regge's $(\eps -n)-$cone
[1]. The $(\eps -n)-$cones are approximatively realized in an
$n$--dimensional
Regge lattice by the $n-$simplexes.

An $(\eps -2)$--cone can be defined as follows:
Take an Euclidean plane and introduce polar coordinates $\rho$, $\xi$,
where points with the same $\rho$ and angles $\xi$ modulo $2\pi$ are
identified. We obtain the $(\eps -2)-$cone by replacing $2\pi$ with
$2\pi-\eps$. Note that the deficit angle $\eps$ is allowed to be negative.
Turning to higher dimensions, we define an $(\eps -n)-$cone
as the direct product $R^{n-2}\times (\eps -2)-$cone, compare Fig.1.
It is convenient to choose
cylindrical coordinates $\rho$, $\xi$, $z_1$,...$z_{n-2}$, thus
$$ds^2=d\rho^2+\rho^2 d\xi^2+d{z_1}^2+...+d{z_{n-2}}^2\,.\eqa$$
\vskip 5truecm
\noindent
{\bf Fig.1:} Definition of Regge's $(\eps -n)-$ cone. The disc with
deficit angle $\eps$ represents the $(\eps -2)-$cone. It is parametrized by
polar coordinates $\rho$, $\xi$. The straight line symbolizes $R^{n-2}$,
which is the subset for $\rho =0$.
\medskip\goodbreak
\noindent
Parallel transport in an $(\eps -n)-$cone is trivial except for
encirceling the $(n-2)$--dimensional flat submanifold $\rho=0$. In order to be
flat, the plane $z_i =$ const, and hence the $(\eps -n)-$cone, is lacking
an angle $\eps$. Therefore parallel transport around the submanifold $\rho =0$
causes a rotation by an angle $\eps$  within this plane,
orthogonal to the submanifold. This is why the curvature
of the $(\eps -n)-$cone resides
at the $(n-2)$--dimensional submanifold $\rho =0$.

The next step is the definition of what we will call a $C^n$--cone. This
$C^n$--cone will turn out to be the isometrically embedded
$(\eps -n)$--cone in $\Reel^{n+1}$ (n=2,3,4).
We concentrate on a positive deficit angle first, $\eps >0$:
Consider $S^n(r)$, the
$n-$sphere of radius $r$ embedded
in $\Reel^{n+1}$. By introducing spherical coordinates $r$, $\p$,
$\t_1$,..,$\t_{n-1}$, it is possible to characterize a
parallel of latitude of the sphere by $\t_i =const$.
We define a specific $C^n$--cone to be spanned by the tangent vectors
$\vec{{\partial \over{\partial\t_i}}}$ at a parallel of latitude of a specific
$n-$sphere:
$$\vec{x}=\vec{p}(\p) +t_1\vec{\partial\over{\partial\t_1}}+ ...
  +t_{n-1}\vec{\partial\over{\partial\t_{n-1}}}\,.\eqa$$
The symbol $\vec{p}(\p)$ denotes an
arbitrary point of the parallel of latitude, the parameters
$t_1,...,t_{n-1}$ are real, and $\vec{x}$ is a point of the
$C^n$--cone. Accordingly a $C^n$--cone is parametrized by $n$ parameters
$\p,t_1,...,t_{n-1}$ and determined by the $n$ quantities
$r,\t_1$, $...$, $\t_{n-1}$ which characterize the parallel of latitude.
The tangent spaces of the $C^n$--cone coincide
at the parallel of latitude with the tangent spaces of the related $n-$sphere.
Hence, $C^n$--cone and $n-$sphere are (to first order) indistinguishable
there. Fig.2 shows the example of a $C^2$--cone.

\vskip 9truecm
\noindent
{\bf Fig.2:} The $C^2-$cone that is related to a specific parallel of
latitude (this determines $\t$) of a specific $2-$sphere
(this determines $r$). It is spanned by the (embedded) tangent vectors
$\vec{\partial \over {\partial \t}}$ of the parallel of latitude.
\medskip\goodbreak

We focus now on four dimensions, $n=4$. This contains the lower
dimensional cases $n=2,3$ as special cases.
The parametrization of the $C^4$--cone is given by
$$\vec{x}=\vec{p}(\p) +t_1\vec{\partial\over{\partial\t_1}}+t_2\vec{\partial
  \over{\partial\t_2}}+t_3\vec{\partial\over{\partial\t_3}}\,,\eqa$$
or explicitly
$$\eqalign{\pmatrix{x_1\cr x_2\cr x_3\cr x_4\cr x_5\cr} &=
r\pmatrix{\cos\p\,\sin\t_1\,\sin\t_2\,\sin\t_3\cr
                   \sin\p\,\sin\t_1\,\sin\t_2\,\sin\t_3\cr
 \cos\t_1\,\sin\t_2\,\sin\t_3\cr\cos\t_2\,\sin\t_3\cr\cos\t_3}+
t_1 r\pmatrix{\cos\p\,\cos\t_1\,
\sin\t_2\,\sin\t_3\cr \sin\p\,\cos\t_1\,\sin\t_2\sin\t_3\cr -\sin\t_1\,
\sin\t_2\sin\t_3\cr 0\cr 0\cr} \cr
&+t_2 r\pmatrix{\cos\p\,\sin\t_1\,
\cos\t_2\,\sin\t_3\cr \sin\p\,\sin\t_1\,\cos\t_2\,\sin\t_3\cr
\cos\t_1\,\cos\t_2\,\sin\t_3\cr
-\sin\t_2\,\sin\t_3\cr 0\cr}
+t_3 r\pmatrix{\cos\p\,\sin\t_1\,\sin\t_2\,\cos\t_3\cr
                   \sin\p\,\sin\t_1\,\sin\t_2\,\cos\t_3\cr
 \cos\t_1\,\sin\t_2\,\cos\t_3\cr\cos\t_2\,\cos\t_3\cr -\sin\t_3\cr}\,.\cr}
\eqa$$
{}From (2.4) we find the metric of the $C^4$--cone which is induced by
the Euclidean metric of $\Reel^5$:
$$ds^2={d_E}^2d\p^2+r^2\sin^2\t_2\,\sin^2\t_3\,d{t_1}^2+r^2
\sin^2\t_3\,d{t_2}^2
+r^2d{t_3}^2\,,\eqa$$
where
$$\eqalign{d_E=r(\sin\t_1\,\sin\t_2\,\sin\t_3 &+t_1\cos\t_1\,\sin\t_2\,
\sin\t_3+t_2\sin\t_1\,\cos\t_2\,\sin\t_3\cr &+t_3\sin\t_1\,\sin\t_2\,
\cos\t_3)\,.\cr}\,\eqa$$
The label $d_E$ denotes
the Euclidean distance between a point of the
$C^4$--cone and the subset $x_1=x_2=0$ in $R^5$.
The ``tip'' of the $C^4$--cone is defined by the equation $d_E=0$.
This, together with (2.4), yields
$$x_1=x_2=0\,,\;x_3\cos\t_1\,\sin\t_2
\,\sin\t_3+x_4\cos\t_2\sin\t_3+x_5\cos\t_3=r\,.\eqa$$
Equation (2.7) represents a parametrization of the two-dimensional tip
in $R^5$. From (2.5) one may derive that the Riemannian curvature of the
$C^4-$cone vanishes, except for the tip where the metric is ill defined.
\bigskip

\noindent
{\bf 2.2. Explicit construction of the isometry}

According to (2.1), the metric of the $(\eps-4)-$cone reads
$$ds^2=\rho^2 d\xi^2 +d\rho^2 + d{z_1}^2+d{z_2}^2\,.\eqa$$
We relate the orthonormal coframes in (2.5) and (2.8)
by a rigid rotation, i.e.
by a constant orthogonal matrix, according to
$$\pmatrix{d_E d\p \cr r\sin\t_2\,\sin\t_3\,dt_1\cr r\sin\t_3\,dt_2\cr rdt_1
\cr}=
\pmatrix{1 & 0 & 0 & 0\cr 0 &e_1 & -e_2 &
-e_3\cr 0 & e_2 & e_1+{{{e_3}^2}\over{{e_2}^2+{e_3}^2}}(1-e_1) &
-{{e_2 e_3}\over{{e_2}^2+{e_3}^2}}(1-e_1)\cr 0& e_3 &
-{{e_2 e_3}\over{{e_2}^2+{e_3}^2}}(1-e_1) &
e_1+{{{e_2}^2}\over{{e_2}^2+{e_3}^2}}(1-e_1)\cr}
\pmatrix{\rho d\xi\cr d\rho\cr dz_1\cr dz_2\cr}\,.\eqa$$
The constants $e_i$ are given by
$$e_1:={{\cos\t_1}\over{\o}}\,,\quad e_2:={{\sin\t_1\,\cos\t_2}\over{\o}}\,,
\quad e_3:={{\sin\t_1\,\sin\t_2\,\cos\t_3}\over{\o}}\,,\eqa$$
$$\o:=\sqrt{1-\sin^2\t_1\,\sin^2\t_2\,\sin^2\t_3}\,.$$
Equation (2.9) defines a coordinate transformation
$$(\p,t_1,t_2,t_3) \longmapsto (\xi,\rho, z_1,z_2)$$
which allows to replace in the parametrization (2.4) of the $C^4-$cone
the coordinates $(\p,t_1,t_2,t_3)$ by $(\xi,\rho, z_1,z_2)$.
The replacement yields a mapping $f$ from the $(\eps-4)-$cone onto the
$C^4-$cone:
$$f:\qquad (\eps -4)- \>{\rm cone}\quad\longmapsto\quad C^4-{\rm cone}$$
$$f(\xi,\rho,z_1,z_2)=
\pmatrix{\rho\o\cos({{\xi}\over\o})\cr
         \rho\o\sin({{\xi}\over\o})\cr
         \rho \sin\t_1\,\sin\t_2\,\sin\t_3\,k_{11}+z_1 k_{12}+z_2 k_{13}\cr
         \rho \sin\t_1\,\sin\t_2\,\sin\t_3\,k_{21}+z_1 k_{22}+z_2 k_{23}\cr
         \rho \sin\t_1\,\sin\t_2\,\sin\t_3\,k_{31}+z_1 k_{32}+z_2 k_{33}\cr}
\,,\eqa$$
where
$$\eqalign{k_{11}&:={{-\cos\t_1\,\sin\t_2\,\sin\t_3}\over\o}\,,\;\;
k_{12}:={{\cos\t_2}\over{\o}}\,,\;\;
k_{13}:={{\sin\t_2\,\cos\t_3}\over{\o}}\,,\cr
k_{21}&:={{-\cos\t_2\,\sin\t_3}\over\o}\,,\;\;
k_{22}:={{-\sin\t_2\,\cos\t_1}\over\o}-{{\cos^2\t_3\,\sin\t_2(1-{{\cos\t_1}
\over{\o}})}\over{\cos^2\t_2\,+\sin^2\t_2\,\cos^2\t_3}}\,,\cr
k_{23}&:={{\cos\t_1\,\cos\t_2\,\cos\t_3}\over{\o}}+{{\cos\t_2\,\cos\t_3
(1-{{\cos\t_1}\over{\o}})}\over{\cos^2\t_2\,+\sin^2\t_2\,\cos^2\t_3}}\,,\;\;
k_{31}:=-{{\cos\t_3}\over\o}\,,\cr
k_{32}&:={{\sin\t_2\,\sin\t_3\,\cos\t_2\,\cos\t_3\,(1-{{\cos\t_1}
\over{\o}})}\over{\cos^2\t_2\,+\sin^2\t_2\,\cos^2\t_3}}\,,\cr
k_{33}&:=-{{\sin\t_3\,\cos\t_1}\over{\o}}-{{\cos^2\t_2\,\sin\t_3(1-{{\cos\t_1}
\over{\o}})}\over{\cos^2\t_2\,+\sin^2\t_2\,\cos^2\t_3}}\,.\cr}\eqa$$

Both cones are equivalent since the rotation in (2.9) just corresponds to
a different choice of an orthonormal coframe. Therefore
the mapping $f$ should be an isometry. We prove this by choosing
$$v:=\rho{\partial\over{\partial\rho}}+\xi{\partial\over{\partial\xi}}
+z_1{\partial\over{\partial z_1}}+z_2{\partial\over{\partial z_2}}\,,\eqa$$
and calculating
$$<\vec{v},\vec{v}>_{R^4}\;=\;\rho^2 +\rho^2\xi^2+{z_1}^2+{z_2}^2\,,\eqa$$
$$<\vec{f_*v},\vec{f_*v}>_{R^5}\;=\;<{{\partial f^i}\over{\partial v^j}}v^j,
{{\partial f^i}\over{\partial v^j}}v^j>_{R^5}\;=\;
\rho^2 +\rho^2\xi^2+{z_1}^2+{z_2}^2\,.\eqa$$
Hence
$$<\vec{v},\vec{v}>_{R^4}\quad = \quad
<\vec{f_* v},\vec{f_* v}>_{R^5}\,,\eqa$$
i.e. $f$ is an isometry, indeed.
\bigskip

\noindent
{\bf 2.3. Negative deficit angles}

So far we dealt exclusively with positive deficit angles. This can be seen
by calculating $\eps$ explicitly: The distance $\varrho$ between the tip
of a $C^4$-cone, given by (2.7), and a parallel of latitude turns out to be
$$\varrho\;=\;r{{\sin\t_1\,\sin\t_2\,\sin\t_3}\over{\sqrt{1-\sin^2\t_1\,
\sin^2\t_2\,\sin^2\t_3}}}\,.\eqa$$
Thus the length of a parallel of latitude is given by
$(2\pi-\eps)\varrho$ or, via the embedding, by
$2\pi r\sin\t_1\,\sin\t_2\,\sin\t_3$. Equating both expressions yields
$$\eps\;=\;2\pi(1-\sqrt{1-\sin^2\t_1\,\sin^2\t_2\,\sin^2\t_3})>0\,.\eqa$$

The embedding of an $(\eps-n)-$cone with a negative deficit angle is
analogous to the case with positive angle.
We replace $S^n(r)$ by $H^n(r)$, which
is one sheet of a two--sheeted hyperboloid:
$$H^n(r)=\{ \vec{x}\in R^{n+1}|<\vec{x},\vec{x}>_{L^{n+1}}:=
-{x_1}^2-{x_2}^2...-{x_n}^2+{x_{n+1}}^2=r^2,\;x_{n+1}>0 \} \eqa$$
The Lorentzian metric $<,>_{L^{n+1}}$ induces a Riemannian metric
on $H^n(r)$. The sheet $H^n(r)$ possesses a constant Riemannian curvature of
amount $-{1\over {r^2}}$ with respect to this metric.

The substitutions $\sin\t_{n-1}\longrightarrow\sinh\t_{n-1}$,
$\cos\t_{n-1}\longrightarrow\cosh\t_{n-1}$ lead from spherical coordinates
on the sphere to hyperbolical coordinates on $H^{n}(r)$. The
isometrically embedded $(\eps-n)-$cone  with negative deficit angle is given
by (2.3), where in this case $\vec{p}(\p)$ denotes
a point of a parallel of latitude $\t_i =$ const. on $H^n(r)$.
This can be proven as in the case with positive deficit angle: We
rewrite $\t_{n-1}$ as $i\,(-i\t_{n-1})$ and use $-i\t_{n-1}$ as a new
complex variable $\tilde\t_{n-1}$. Then $\sinh{\t_{n-1}}\,
=i\sin\tilde\t_{n-1}$ and $\cosh{\t_{n-1}} = \cos\tilde\t_{n-1}$. This leads
exactly to the former starting point (2.5).
The factor $i$ in front of $\sin\tilde\t_{n-1}$ takes into account the
Lorentzian metric of $R^{n+1}$.
Computation of the distance $\varrho$ between the tip of a cone with negative
deficit angle and a related parallel of latitude yields
$$\varrho \;=\;r{{\sin\t_1\,\sin\t_2\,\sinh\t_3}\over
{\sqrt{1+\sin^2\t_1\,\sin^2\t_2\,\sinh^2\t_3}}}\,,\eqa$$
and we obtain
$$\eps \;=\;
2\pi\Bigl(1-\sqrt{1+\sin^2\t_1\,\sin^2\t_2\,\sinh^2\t_3}\Bigr)<0\,.
\eqa$$

Calculation of $\eps$ by means of the parallel transport equation related
to the Riemannian connection of $S^n(r)$ or $H^n(r)$ yields, of course,
the same result.

\sectio{\bf General parallel transport on a Regge lattice}

The notion of parallel transport on a Regge lattice becomes transparent now:

Riemannian parallel transport on a Regge
lattice is trivial except
for surrounding hinges with nonzero deficit angle. The result of parallel
transport around such a hinge is independent of the path chosen.
Thus we may choose a parallel of latitude of the related
$(\eps-n)-$cone and identify this, via the isometrical embedding (2.11),
with a parallel of latitude of a specific space of constant
curvature.

In order to relate a specific hinge of the Regge lattice (with adjacent
$n-$simplexes) to a certain $(\eps-n)-$cone and, in turn, to a
parallel of latitude of a space of constant curvature,
it is required to read off the values of $\eps$ and $\rho$ from the
lattice. Then we obtain for a positive deficit angle by (2.17) and (2.18)
values for $r$ and $(\sin\t_1\,\sin\t_2\,\sin\t_3)$, while for a negative
deficit angle we use (2.20) and (2.21) to obtain values for $r$ and
$(\sin\t_1\,\sin\t_2\,\sinh\t_3)$.
The deficit angle $\eps$ can be read
off exactly, but how do we read off $\rho$? The distance $\rho$ between the
tip of a $(\eps -n)$--cone and a parallel of latitude corresponds in a Regge
lattice to the distance between a hinge and a path that encircels the hinge.
If we restrict ourselves to define paths in a Regge lattice exclusively
on a corresponding dual lattice, we may assign to $\rho$ an average
distance between a hinge and a path that encircels it.
Adopting, for example, a barycentric dual lattice, the distance $\rho$
can be defined by the average of the Euclidean distances between the
hinge and the barycenters of adjacent n--simplexes. This procedure is the
only approximation that enters the transition from the path surrounding
a hinge on a Regge lattice to a parallel of latitude of $S^n(r)$ or $H^n(r)$
and expresses the approximative character of the Regge lattice. The
transition requires the metric of the Regge lattice
since $\eps$ and $\rho$ are functions of the link lengths.

It turned out in Sec.2
that the $(\eps-n)-$cone is build from the tangent spaces
of a parallel of latitude of a space of constant curvature. The $n-$simplexes
related to a hinge locally approximate some $(\eps-n)-$cone in a sense that
each of them constitutes (a part of) the tangent space at one point of
the ``underlying'' parallel of latitude. Therefore each $n-$simplex represents
a tangent space of the manifold that is approximated by the Regge lattice.
We infer from this that a general
parallel transport in a Regge lattice is provided by a
prescription which connects adjacent $n$--simplexes, i.e. neighbouring
tangent spaces. Such a prescription can be derived immediately by
regarding the  intuitive and illuminating description of parallel transport
in a manifold which is due to {\'E}.Cartan [11]: Let $p$ and $p'$ be two
infinitesimally close points of a manifold $M$, their affine tangent spaces
are denoted by $T_{p}M$ and $T_{p'}M$, both are equipped with a vector basis
$e_{\alpha}$ and ${e'}_{\alpha}$, respectively. The affine tangent spaces
$T_pM$ and $T_{p'}M$ are compared by means of an affine (Cartan-)connection
$(\vta^\alpha,\Gamma_{\alpha}{}^{\beta})$  as follows:
$$p'  = p+\vartheta^{\alpha}(p')\,e_{\alpha}\,,\eqa$$
$$ {e'}_{\alpha}=\Bigl({\Gamma}_{\alpha}{}^{\beta}(p') +
\delta_{\alpha}^{\beta}\Bigr)\,e_{\beta}{}\,.\eqa$$
Equations (3.1) and (3.2) define an affine transformation from
$T_pM$ to $T_{p'}M$ and
specify parallel transport from $p$ to $p'$. Integration of the one forms
$\vta^\a$ and $\Gamma_{\a}{}^\b$ around infinitesimally closed loops
yields the Cartan structure equations
$$ T^{\alpha}:=d\vartheta^{\alpha} + \vartheta^{\beta}\wedge{\Gamma}_{\beta}
  {}^{\alpha}\,,\eqa$$
$$R_{\alpha}{}^{\beta}:=d{\Gamma}_{\alpha}{}^{\beta} +
   {\Gamma}_{\alpha}{}^{\gamma}\wedge{\Gamma}_{\gamma}{}^{\beta}\,.\eqa$$
Therefore sucessive affine transformations around an infinitesimally
closed loop can be replaced by one affine transformation, which is determined
by the two--forms torsion $T^\a$ and curvature $R_\a{}^\b$.
The torsion determines the {\it translational} part, whereas the curvature
determines the {\it linear} part of the resulting affine transformation.
We may detect whether the linear part is orthogonal or not if
we equip $M$ with a metric $g$. Then it is possible to express
shears and/or dilations in terms of the nonmetricity one-form [12]:
$$Q_{\a\b}:=-Dg_{\a\b}=-dg_{\a\b}+\Gamma_{\a}{}^\gamma g_{\gamma\b}+
\Gamma_{\b}{}^{\gamma}g_{\a\gamma}\,.\eqa$$

This implies that a general parallel parallel transport in a Regge lattice
is defined by affine transformations between adjacent $n-$simplexes.
A set of affine transformations, one affine transformation for each pair
of adjacent $n-$simplexes, may be called a discrete affine connection.
The approximative
character of the Regge lattice is expressed by the fact that the points
$p$ and $p'$ are no longer infinitesimally close. Parallel transport around
a hinge is performed by successive affine transformation while passing from
one simplex to another. Curvature, torsion and nonmetricity
associated with a hinge
are defined by the affine transformation which results from
this discrete parallel
transport around the hinge.

To become more explicit, we consider some closed loop in a Regge lattice
that surrounds one hinge by passing through $k$  different
$n$--simplexes $S_1,\dots ,S_k$.
Parallel transport is specified by a discrete affine connection
$(\vartheta^\a_{\{lm\}},\Gamma_{\{lm\}\a}{}^\b)$, where the affine
transformation
connecting the adjacent $n$--simplexes $l$ and $m$ is given by
$\vartheta^\a_{\{lm\}}$ and $\Gamma_{\{lm\}\a}{}^\b$. Each simplex
represents
an affine tangent space so that we choose an origin $P$ and
a frame $e_\a$ in each of them. Parallel transport from simplex $S_1$
to $S_2$ is specified by (cf. (3.1), (3.2))
$$
p_{\{2\}} = p_{\{1\}} + \vartheta^\a_{\{12\}}e_{\{1\}\a}\,,\eqa
$$
$$\eqalign{e_{\{2\}\a} &= (\Gamma_{\{12\}\a}{}^\b+\d_\a^\b)e_{\{1\}\b} \cr
                     &=: \Lambda_{\{12\}\a}{}^\b e_{\{1\}\b}\,,\cr} \eqa
$$
or, in a shorter notation, by $(\vartheta_{\{12\}}^\a ,
\Lambda_{\{12\}\a}{}^\b)_{S_1\rightarrow S_2}$. Similarly, we have for parallel
transport from $S_2$ to $S_3$ the expression
$(\vartheta_{\{23\}}^\a , \Lambda_{\{23\}\a}{}^\b)_{S_2\rightarrow S_3}$
such that parallel transport from $S_1$ to $S_3$ is described by
$$
(\vartheta_{\{23\}}^\a +\vartheta_{\{12\}}^\a
\Lambda_{\{23\}\a}{}^\b , \Lambda_{\{23\}\a}{}^\c \Lambda_{\{12\}\c}{}^\b)_{S_1
\rightarrow S_3} =:
(\vartheta_{\{13\}}^\a , \Lambda_{\{13\}\a}{}^\b)_{S_1\rightarrow S_3}\,.\eqa
$$
Proceeding with parallel transport to $S_k$ yields
$$\eqalign{
(\vartheta_{\{(k-1)k\}}^\a +\vartheta_{\{1(k-1)\}}^\a
\Lambda_{\{(k-1)k\}\a}{}^\b , \Lambda_{\{(k-1)k\}\a}{}^\c
\Lambda_{\{1(k-1)\}\c}{}^\b)_{S_1\rightarrow S_k}&\cr  =:
(\vartheta_{\{1k\}}^\a , \Lambda_{\{1k\}\a}{}^\b)_{S_1\rightarrow S_k}&\,,
\cr}\eqa
$$
and we obtain the affine transformation that results from parallel
transport around this loop,
$$
(\vartheta_{\{k1\}}^\a +\vartheta_{\{1k\}}^\a
\Lambda_{\{k1\}\a}{}^\b , \Lambda_{\{k1\}\a}{}^\c
\Lambda_{\{1k\}\c}{}^\b)_{S_1\rightarrow S_1} =:
(T^\a, R_\a{}^\b + \d_\a^\b)_{S_1\rightarrow S_1}\,,\eqa
$$
where $T^\a$ and $R_\a{}^\b$ denote torsion and curvature associated to this
loop. We note the recursive definition of $\vartheta^\a_{\{1l\}}$ and
$\Lambda_{\{1l\}\a}{}^\b$ for $l\in \{3,...,k\}$.
Equation (3.10) constitutes the discrete equivalent to (3.3) and (3.4).
The nonmetricity can be calculated by using a metric $g$ in $S_1$:
$$
Q_{\a\b}\; = \; -g(R_\a{}^\c e_\c, R_\b{}^\c e_\c)\,.\eqa
$$
The definition of torsion and curvature in both the Cartan structure equations
(3.3), (3.4) and their discrete analogue (3.10) is completely independent of
metric and link lengths, respectively.
The link lengths determine the manifold approximated by the Regge lattice,
while the discrete affine connection approximates a general parallel transport
on this manifold.
It is just the special case of Riemannian parallel transport
that can be deduced from the link lengths via the deficit angle.

\sectio{\bf Summary}

The isometry (2.11) proves that every Regge lattice locally,
or ``hingewise'', approximates
a space of constant curvature. The explicit identification
between a hinge of the Regge lattice and a specific space of constant
curvature is due to the metric, i.e. the link lengths, of the Regge lattice.
The $n-$simplexes represent affine tangent spaces of the manifold
approximated. Parallel transport in a Regge lattice is a prescription that
maps neighbouring $n-$simplexes onto each other. A priori this prescription
is not determined but may be defined by a discrete
affine connection. Such a discrete affine connection is independent of
the metric of the Regge lattice. This is, of course, analogous to
ordinary differential geometry: The metric alone cannot determine an arbitrary
affine connection. Therefore it seems that attempts to build a non-Riemannian
Regge calculus by relying exclusively on the metric are not promising.

\bigskip
\noindent
{\bf Acknowledgments}

The author is indebted to F.W. Hehl for valuable comments. This work was
supported by a grant of the German Academic Exchange Service
(DAAD-Doktoran\-den\-stipendium aus Mitteln des zweiten
Hochschulsonderprogramms).

\bigskip
\bigskip
\vfill
\eject

\noindent
{\bf References}
\bigskip

\noindent
[1] T.Regge, {\it Nuovo Cim.} {\bf 19}, 558 (1961)

\noindent
[2] R.M.Williams and P.A.Tuckey, {\it Class. Quantum Grav.}
{\bf 9}, 1409 (1992)

\noindent
[3] F.W.Hehl, J.D.McCrea, E.W.Mielke, and Y.Ne'eman, {\it Phys. Rep.}
1995 (in press)

\noindent
[4] P.van Nieuwenhuizen, {\it Phys. Rep.} {\bf 68}, 189 (1981)

\noindent
[5] J.W.Barret, {\it Class. Quantum Grav.} {\bf 4}, 1565 (1987)

\noindent
[6] I.T.Drummond, {\it Nucl. Phys.} {\bf B273}, 125 (1986)

\noindent
[7] V.Khatsymovsky, {\it Class. Quantum Grav.} {\bf 6}, L249 (1989)

\noindent
[8] A.Kheyfets, N.J. LaFave, and W.A.Miller, {\it Phys. Rev.}
{\bf D39}, 1097 (1989)

\noindent
[9] C.Holm and J.D.Hennig, in {\it Group theoretical methods in physics},
Lecture Notes in Physics {\bf 382} (Springer, 1991), p.556

\noindent
[10] M.Caselle, A.D'Adda, and L.Magnea, {\it Phys. Lett.} {\bf B232},
457 (1989)

\noindent
[11] {\'E}.Cartan, {\it On Manifolds with an Affine Connection and the
Theory of General Relativity}, English translation of the French original
(Bibliopolis, Napoli 1986)

\noindent
[12]  J.A.Schouten, {\it Ricci Calculus}, 2nd ed. (Springer, Berlin 1954)

\vfill
\bye